\documentclass{PoS}
\usepackage{graphicx}
\usepackage[lofdepth,lotdepth]{subfig}
\usepackage[rightcaption]{sidecap}
\graphicspath{ {images/} }
\usepackage{lineno}

\title{Moli\`{e}re radius measurement using a compact prototype of LumiCal in a test set-up}

\ShortTitle{Moli\`{e}re radius measurement using a compact prototype of LumiCal in a test set-up}

\author{\speaker{Veta Ghenescu}\thanks{on behalf of the FCAL Collaboration}\\
        Institute of Space Science,\\
Atomistilor 409 street, Magurele, Romania\\
        E-mail: \email{ghenescu@spacescience.ro}}
				
\abstract{The FCAL collaboration has performed a design study for luminometers at future electron-positron colliders. Compact sampling calorimeters with precisely positioned silicon sensors and a fast readout will reach the necessary performance even in the presence of background from beamstrahlung and two-photon processes. A prototype calorimeter has been built with special focus on ultra-thin fully instrumented sensor planes to ensure a very small effective Moli\`{e}re radius. Preliminary results of measurements in a 5 GeV electron beam are presented.\
}

\FullConference{The 39th International Conference on High Energy Physics (ICHEP2018)\\
		4-11 July, 2018\\
		Seoul, Korea}

\begin{document}

\section{Introduction}
Two concepts of future e$^+$e$^-$  linear collider are currently considered, the ILC~\cite{ilc1} and CLIC~\cite{clic2}. In all detectors at small polar angles, a calorimeter for precision measurements, the LumiCal, and a radiation hard calorimeter, the BeamCal, are foreseen. 
These extend the calorimeter solid-angle coverage to almost 4$\pi$. The LumiCal is designed to measure the luminosity with a precision of better than 10$^-$$^3$ at 500 GeV centre-of-mass energy and 3x10$^-$$^3$ at 1 TeV centre-of-mass energy at the ILC, and with a precision of 10$^-$$^2$ at CLIC up to 3 TeV. The BeamCal will tag electrons and positrons that are only slightly deflected in peripheral scattering events, will perform a bunch-by-bunch estimate of the luminosity and, supplemented by a pair monitor, will assist beam tuning when included in a fast feedback system. 
The LumiCal and BeamCal are designed as cylindrical sensor-tungsten sandwich electromagnetic calorimeters. In the ILC detector design, both consist of 30 tungsten layers of 3.5 mm thickness, each corresponding to one radiation length, interspersed with sensor layers placed in the one millimeter gap between absorber plates. One of the stringent requirements for the calorimeter prototype is the compactness. For LumiCal, silicon pad sensors are envisaged. For BeamCal, due to its lower polar-angle range, high radiation-tolerance is required and therefor, GaAs sensors are under development. Other sensor technologies, including silicon diodes, sapphire and silicon carbide, are also under consideration.  
In this paper, results from tests in an electron beam of the first LumiCal prototype with ultra-thin sensor layers are reported. The design of the thin sensor layers and the experimental set-up at the test beam are described. Preliminary results of the transverse electromagnetic shower development parametrized as an effective Moli\`{e}re radius of this very compact LumiCal prototype are presented.

\section{Design of thin LumiCal sensor layers}

The first LumiCal compact prototype was successfully built and tested in a multilayer configuration at the DESY-II with a beam of secondary electrons with energies between 1 and 6 GeV. The fully assembled sensor layer prototype, developed for this study, achieved a total thickness of 650$\mu$m. 
The LumiCal sensor~\cite{silicon4}  is made from a \textit{N-type} silicon, with a thickness of 320$\mu$m. It is shaped as a ring segment of 30$^\circ$ subdivided into four sectors. Each sector is segmented in the radial direction with 64 pads of 1.8 mm pitch. The LumiCal silicon sensor was glued with epoxy to a 120$\mu$m thick front-end board made of flexible Kapton-copper foil, and then ultrasonic wire bonding was used to connect conductive traces on the fan-out to the sensor pads.  The high voltage was supplied by a 70$\mu$m thick Kapton-copper foil, glued  on the back side of the sensor with conductive glue. In order to obtain the necessary mechanical stability the assembled sensor is embedded in a carbon fibre envelope forming a detector plane. The total thickness of the envelope is about 650$\mu$m, with the part supporting the sensor thinned down to about 120$\mu$m. 

\section{Test beam set-up}
The LumiCal prototype has been operated in test beam campaigns in 2015 and 2016 at DESY in an electron beam of energies between 1 and 5 GeV. Eight ultra-thin detector planes were assembled for the 2016 test beam including one assembled using TAB technology~\cite{tab_5}.  

\begin{figure}[h]
 \centering
  \includegraphics[width=0.84\textwidth]{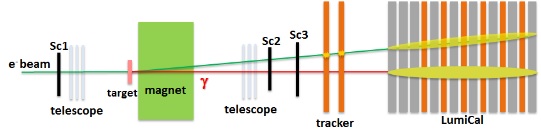}
   \caption{\textit{Test-beam setup, simplified view (not to scale).}}
  \label{fig:desy_setup}
\end{figure}

\begin{figure}[!h]
   \begin{minipage}{0.45\textwidth}
     \centering
     \includegraphics[width=.78\linewidth]{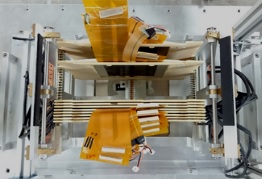}
     \caption{\textit{Top view on the assembled LumiCal prototype}\label{fig:lumical_fig4}}\hspace*{\fill}
   \end{minipage}\hfill
   \begin{minipage}{0.5\textwidth}
     \centering
     \includegraphics[width=.66\linewidth]{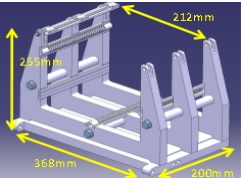}
     \caption{\textit{Dimensions of the mechanical frame for the positioning of detector and absorber planes.}}\hspace*{\fill}\label{mech_frame}
   \end{minipage}\hfill
\end{figure}

A schematic view of the DESY setup is presented in Figure~\ref{fig:desy_setup}. Bremsstrahlung photons are produced by the electron beam hitting a copper target installed upstream close to a dipole magnet. The magnetic field was chosen to allow both photons and electrons to travel within the acceptance of the second telescope arm and to hit LumiCal with a separation large enough to be resolved by the calorimeter. For triggering such events with low energy photons, all scintillator counters (Sc1, Sc2 and Sc3) are used in coincidence. A telescope composed of six layers of MIMOSA-26 chips arranged in two arms of three sensors on each side of the magnet aided the track reconstruction. Two detector planes are used as $"$tracker $"$ as shown in Figure~\ref{fig:desy_setup} and six separated by one absorber tungsten layer as shown in Figures~\ref{fig:desy_setup} and ~\ref{fig:lumical_fig4}.  A crucial requirement of the measurements was to connect all 256 pads from each sensor layer to the front-end board. In order to be able to read out all pads of a sensor the APV25 chip~\cite{apv25_7} was chosen. A mechanical structure~\cite{mech_frame8} (Figure~\ref{mech_frame}) allows for the installation of the tungsten absorber layers and the detector planes with a precision better than 50$\mu$m.   

\section{Results}
The focus of the on-going analysis of the test beam data is on electromagnetic shower development in the longitudinal and transverse direction, as well as
 on the electron/photon separation. 

\begin{SCfigure}[1.0][!h]
 \centering
	\caption{\textit{The transverse shower profile $<$E$_m$$>$, as a function of d$_{core}$ in units of pads, from beam-test data and the MC simulation, after symmetry corrections and fit. The lower part of the figure shows the ratio of the distributions to the fitted function, for the data (blue) and the MC (red).}}
	\includegraphics[width=0.46\textwidth]{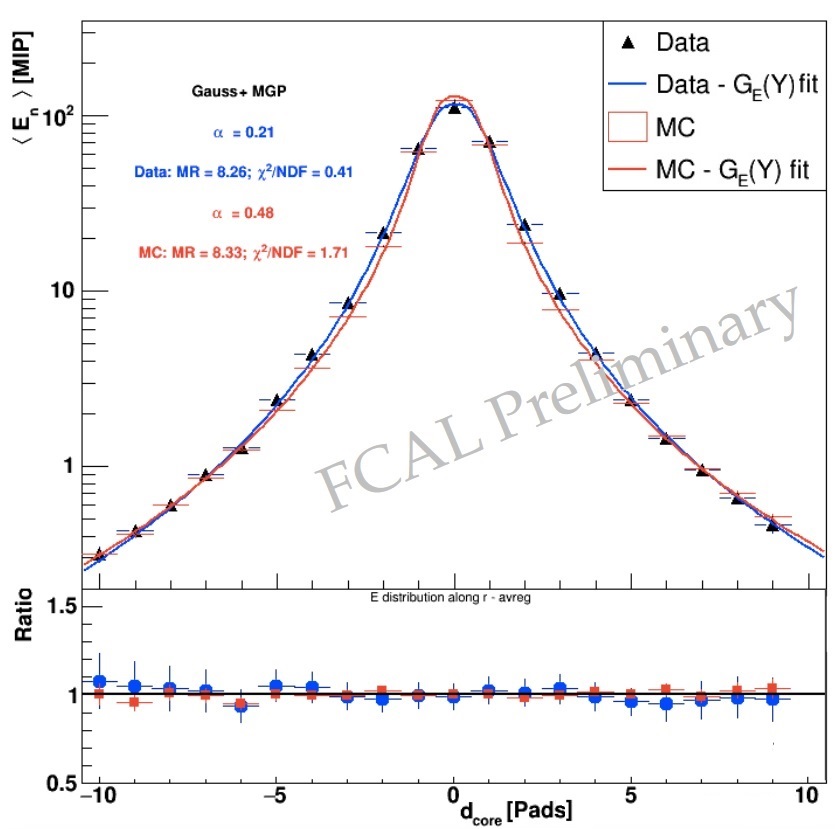}
		\label{shower_transv_6}
\end{SCfigure}

As an example, in Figure~\ref{shower_transv_6}, the preliminary result on the transverse shower profile for an electron energy of 5 GeV is shown. The different steps of the analysis are described in detail in Ref. ~\cite{molier_9}. The measurements are in good agreement with Monte Carlo simulations: the preliminary result obtained for the effective Moli\`{e}re radius as defined in the paramatrisation given in Ref.~\cite{molier_9}  is R$_M$ = 8.26 mm for experimental data and R$_M$= 8.33 mm for the simulation.

\section{Conclusions}
The first LumiCal compact  module prototype was prepared and tested in an electron beam. Using a new technology to built detector planes, the FCAL collaboration succeeded to develop, produce and operate detector planes for LumiCal with a thicknessof about 650 $\mu$m. This allows for the construction of a calorimeter prototype of an unprecedented compactness. The transverse electromagnetic shower shape was investigated, and the effective Moli\`{e}re radius for 5 GeV electrons was found to be 8.3mm, in good agreement with Monte Carlo simulations.

\section*{Acknowledgement}
	This activity was partially supported by the Romanian Space Agency, R$\&$D projects 160/2017, 168/2017 and by the EU H2020 R\&I programme under Grant No. 654168 (AIDA-2020).

\end{document}